**Machine Learning Design of Perovskite Catalytic Properties**

**Authors:** Ryan Jacobs[1,*], Jian Liu[2], Harry Abernathy[2], Dane Morgan[1]

[1] Department of Materials Science and Engineering, University of Wisconsin-Madison, Madison, WI, 53706, USA.

[2] National Energy Technology Lab, Morgantown, WV, 26505, USA.

*Corresponding author e-mail: rjacobs3@wisc.edu

# Abstract

Discovering new materials that efficiently catalyze the oxygen reduction and evolution reactions is critical for facilitating the widespread adoption of solid oxide fuel cell and electrolyzer (SOFC/SOEC) technologies. Here, we develop machine learning (ML) models to predict perovskite catalytic properties critical for SOFC/SOEC applications, including oxygen surface exchange, oxygen diffusivity, and area specific resistance (ASR). The models are based on trivial-to-calculate elemental features and are more accurate and dramatically faster than the best models based on *ab initio*-derived features, potentially eliminating the need for *ab initio* calculations in descriptor-based screening. Our model of ASR enables temperature-dependent predictions, has well calibrated uncertainty estimates and online accessibility. Use of temporal cross-validation reveals our model to be effective at discovering new promising materials prior to their initial discovery, demonstrating our model can make meaningful predictions. Using the SHapley Additive ExPlanations (SHAP) approach, we provide detailed discussion of different approaches of model featurization for ML property prediction. Finally, we use our model to screen more than 19 million perovskites to develop a list of promising cheap, earth-abundant, stable, and high performing materials, and find some top materials contain mixtures of less-explored elements (e.g., K, Bi, Y, Ni, Cu) worth exploring in more detail.

# Main

A key impediment to more widespread adoption of solid oxide fuel and electrolyzer cell (SOFC/SOEC), including reversible (r-SOFC) and proton ceramic fuel cell (PCFC)[1–12] technologies is the availability of electrode materials which are cheap, stable, and can effectively catalyze the oxygen reduction (ORR, for fuel cells) and evolution (OER, for electrolyzers) reactions at reduced operating temperatures of about 500 °C or even lower.[4,5,13] Perovskite oxides are the most popular and well-studied non-precious metal ORR/OER catalysts for current and next generation SOFC/SOEC technologies. Computational discovery of new perovskite electrodes has traditionally centered on the use of first-principles based descriptors of catalytic activity, such as the O p-band center descriptor obtained from density functional theory (DFT) calculations.[14] The O p-band center has been successfully used to form correlations with myriad perovskite properties ranging from oxygen surface exchange rates to electronic work function.[14–24] However, even the use of descriptors like the O p-band center rely on modestly expensive DFT calculations which places constraints on the speed with which one can propose new materials or understand trends in materials properties, while the use of data-driven machine learning (ML) approaches provides a promising avenue for accelerating both understanding and discovery of new promising materials.

The use of ML approaches in materials science has seen a meteoric rise in recent years.[25–29] However, the prediction of perovskite catalytic properties with ML remains in the nascent stages,[30–33] with only a handful of papers using ML to predict properties like ASR and oxygen conductivity.[34–37] The O p-band center property correlations mentioned above can be considered a primitive ML model, where the model has a single feature, the O p-band center, and the model type is often a basic univariate linear regressor. It is likely that more sophisticated data-driven techniques can be utilized for understanding and discovering new high-performing SOFC/SOEC materials. Recently, Xin highlighted the opportunity that ML techniques have in advancing the state-of-the-art of general catalyst design.[38] More specifically for SOFC/SOECs, recently Zhai et al.[34] used a database of 85 area specific resistance (ASR) values, a neural network ML model, and elemental features together with a newly-proposed ionic Lewis acid descriptor, to screen for promising new perovskite materials. Zhai et al. enumerated a set of 6871 possible compositions and selected four promising materials based on low predicted ASR values, and experimentally confirmed these materials are indeed high performing, with ASR values on par or lower than the benchmark high performing material $Ba_{0.5}Sr_{0.5}Co_{0.8}Fe_{0.2}O_3$ (BSCF). In addition, recent work from Xu et al.[35] used a weighted voting regression approach combining multiple ML models to predict the oxygen conductivity in perovskite oxide electrolytes and suggest a number of doped gallates as promising electrolytes.

In this work, we take a purely data-centric ML approach to predicting perovskite catalytic properties and have three major results. The first major result is the development of ML models for perovskite catalytic property prediction and the demonstration of its superiority to the O p-band center descriptor approach for such predictions. We show that ML models based on elemental features can deliver better average accuracy in predicting key properties than O p-band center correlations, at least within our test set, and can be used to make predictions many orders of magnitude faster than O p-band center correlations as the ML requires no DFT calculation. The second key result of this work is the development and assessment of an ML model for ASR predictions. We show that our ASR model can realize low errors through a novel featurization scheme using a combination of elemental features, one-hot encoding of electrolyte type, and a separate ML model predicting the ASR activation energy barrier, which value is then used as a feature in the ASR prediction model. The inclusion of ASR energy barrier enables temperature-dependent ASR predictions. Further, our ASR model shows an ability to predict future promising materials based on temporal cross validation and provides calibrated uncertainty estimates (i.e., error bars). The third and final major result of this work is the application of the ASR ML model to search a large space of approximately 19 million (19M) perovskite compositions and identify new promising cheap, earth-abundant, stable and highly catalytically active perovskite materials.

## Machine learning replaces electronic structure descriptors for catalytic property predictions

We use the database of perovskite catalytic properties from the work of Jacobs et al.[39] This database consists of 749 data points spanning 299 unique perovskite compositions. The data consists of oxygen surface exchange, diffusivity, and ASR values. In our previous work and similar descriptor-based studies, the O p-band performance are full fits to all values in a database. In ML studies, it is more common to assess model performance through cross-validation (CV), where certain subsets of the data are used to train the model, while the remaining data is held out to test the model performance (e.g., train on 80% and test on 20% of the data). In this section, we construct random forest ML models for each catalytic property at T = 500 °C and use random 5-fold CV to assess model errors. All ML model fits and evaluations were performed using the MAterials Simulation Toolkit for Machine Learning (MAST-ML) package.[40] All models use combinations of elemental properties as their features, which are nearly instantaneous to determine for a new compound. More details of the ML model, features, and fitting are given in **Section S1** of the **Supplementary Information (SI)**. To compare to the performance of O p-band center discussed in our previous work, we also assess model errors of a linear model fit to the O p-band center using random 5-fold CV.

**Figure 1** provides a summary of ML model assessment for each catalytic property at T = 500 °C, with comparisons to the linear model fit to the O p-band center. The corresponding plot for T = 800 °C is shown in **Section S2** of the **SI** and shows the same qualitative conclusions. In **Figure 1**, the bars (error bars) denote the average (standard error in the mean) MAE over 25 splits of 5-fold CV. The blue bars are average MAE values for the case where the DFT-calculated O p-band center is used as a single feature with a linear regression model. The green bars are average MAE values for the case of using trivial-to-calculate elemental features with the ML model. The percentage reductions in average MAE by using ML elemental features compared to O p-band center and linear regression are 34.7%, 29.1%, -6.6%, -1.5%, and 27.0% for $k^*$, $D^*$, $k_{chem}$, $D_{chem}$, and ASR, respectively. Therefore, for all properties, the average MAE is within the uncertainty or reduced for the case of ML with elemental features compared to O p-band center with linear regression. This analysis demonstrates that even with limited dataset sizes of order 50 data points, ML can produce prediction errors on par or lower than those obtained using a physically-motivated descriptor such as the O p-band center. In addition, the use of ML here has the advantage that predicting properties of new materials is orders of magnitude faster than performing DFT calculations to obtain the O p-band center or conducting experiments.

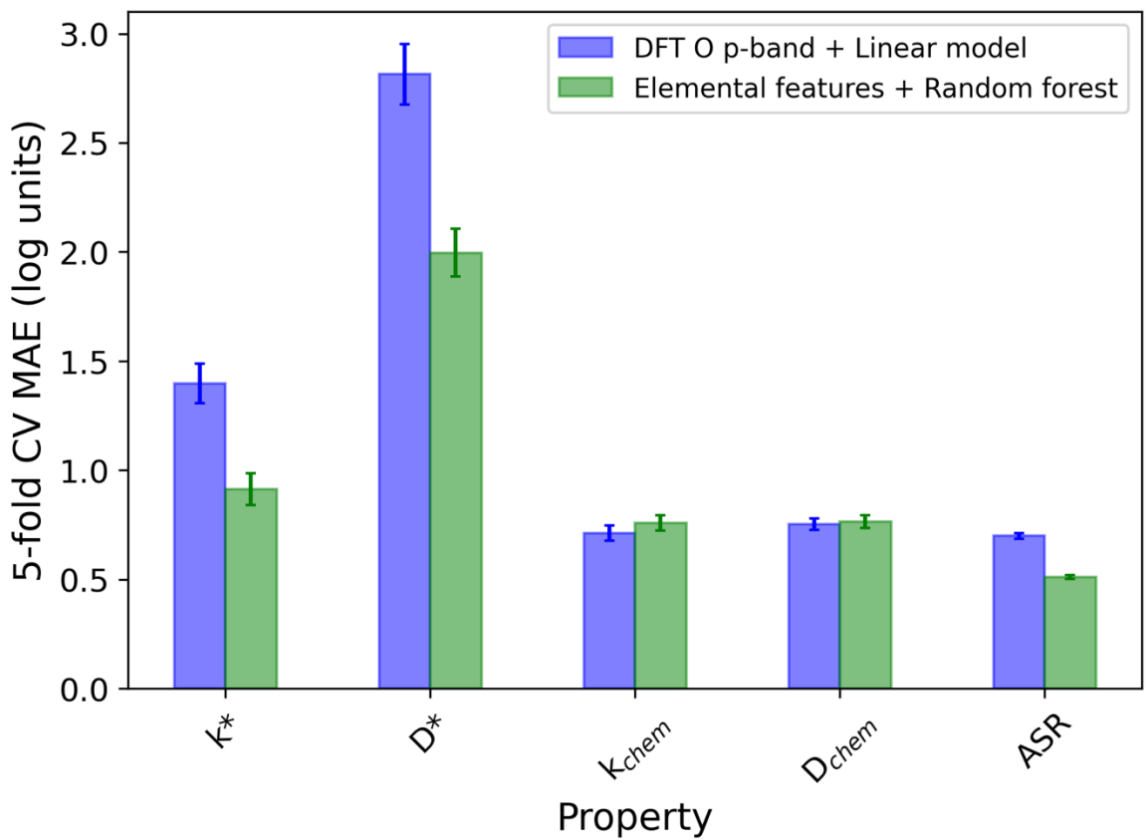

**Figure 1.** Machine learning model random cross validation assessment comparing performance of the DFT-calculated O p-band center descriptor with a linear model and a random forest model using elemental features. These assessments are for T = 500 °C. The units of k* and $k_{chem}$ are cm/s, the units of D* and $D_{chem}$ are cm$^2$/s, and the units of ASR are Ohm-cm$^2$. The error bars are standard errors in the mean of the calculated MAE over 25 splits of 5-fold CV. The ML models of k*, D*, $k_{chem}$ and $D_{chem}$ use only elemental features, while the ML model of ASR uses elemental features and one-hot encoding of the electrolyte type.

## ML model of ASR with low errors, well-calibrated uncertainties, and effective time-dependent materials predictions

**Figure 2** details our ML model performance for predicting ASR. We found that the most effective ASR model is featurized using elemental features, one-hot encoding of electrolyte type, and a feature consisting of a separate ML model prediction of the Arrhenius energy barrier for ASR, which itself uses only elemental features and one-hot encoding of electrolyte type (see **Section S1** of the **SI** for more details). **Figure 2A** and **Figure 2B** contain parity plots of a random forest ML model fit to all the data (full fit) and assessed by random 5-fold CV, respectively. In these parity plots, the blue points are materials which contain 4 or fewer independent experimental measurements, and the green points are designated "well-studied" materials with greater than 4 measurements. The separation of well-studied materials is done because our previous work[39] showed that the properties for these materials are much more amenable to fitting, likely due to their having reduced noise from averaging the multiple measurements. The listed metrics in black text are taken over all the data, while the metrics listed in green text are for the subset of well-studied materials only. The error bars on the points in the parity plots are calibrated uncertainty estimates from the ML predictions. Because our random forest model is an ensemble of decision trees, we can obtain an uncertainty on each prediction by calculating the standard deviation of the predictions of the individual trees. This approach provides a simple ensemble estimate of the prediction uncertainty, but one cannot tell *a priori* if this uncertainty estimate itself is accurate. We follow the approach of Palmer et al.[41] to develop calibrated uncertainty estimates and demonstrate that these calibrated uncertainty estimates are quite accurate. More information on the error bar assessment and calibration is in **Section S3** of the **SI**.

From our full fit model in **Figure 2A**, our ML model is able to accurately fit our ASR database, and the calibrated error bars tend to intersect the *y* = *x* line, which represents perfect prediction. From our 5-fold CV results in **Figure 2B**, the quantitative prediction quality is reduced compared to the full fit, with some amount of overestimation (underestimation) of the true ASR values for the lowest (highest) ASR values in our database. The overestimation of ASR at the lowest range of values does not present an issue from the standpoint of materials screening, as it suggests the ASR predictions may be conservatively higher than the resulting true values, thus minimizing the likelihood of undesirable false positive predictions. It is worth noting that the ML prediction metrics for the subset of well-studied materials are significantly improved compared to predicting on all of the data. This result is consistent with our previous observation of O p-band center trends in from our previous work and suggests that the ML model fit is sensitive to the quality and noise level of the experimental data used.[39]

In **Figure 2C**, we consider the problem of using our ML regression model for the purpose of classifying whether a particular material is expected to have a log ASR value below a given threshold. Using the 5-fold CV results in **Figure 2B**, we find that our ML regression model can correctly find 92% of materials that should have a log ASR < 1 Ohm-cm$^2$ at 500 °C, which represents a relatively high performing material. This result is encouraging as we will use our ML model to screen for new promising perovskite catalysts. Going beyond basic 5-fold cross validation, our model also shows effectiveness at predicting future promising materials. This is shown through temporal cross validation in **Figure 2D**. In **Figure 2D,** the classification metrics denote the ability of the model to correctly classify materials as having log ASR < 1 Ohm-cm$^2$, where the training sets are tranches of materials grouped by year of their initial study, and the corresponding test sets are all materials occurring after the most recent year in the training data. For example, the data point for "1998-2008" means the training data are all materials first studied from 1998-2008, and the test data are all materials first studied from 2009-2021. Note that all future instances of a material initially studied in a given year are placed in the tranche corresponding to the initial year of study, so repeat compositions are not present in both the train and test sets. The results of **Figure 2D** show that our model reaches an F1 classification score of about 0.8 for predicting materials first discovered between 2009-2021 after only training up to materials first discovered by 2008, corresponding to a training set of just 76 data points (52 unique compositions). We find similar results for an ASR model trained at 800 °C with a log ASR classification threshold of -1 Ohm-cm$^2$ (see **Section S2** of the **SI**). We note that using our 800 °C ASR model trained on only the 23 data points (14 unique compositions) from the first tranche of 1998-2003 and predicting materials that are cheap (< $10/kg) and reasonably stable (< 125 meV/atom) (see discussion below for more details) suggests promising compositions in the Ba-Sr-Fe-Co space, indicating this initial model could have been used to predict materials similar to BSCF to be promising prior to its initial discovery in 2004. It is impressive to note that this model also suggests that doping Zr on the B-site should result in stable and highly active materials, long before the initial publication of BFCZ20 in the year 2013.

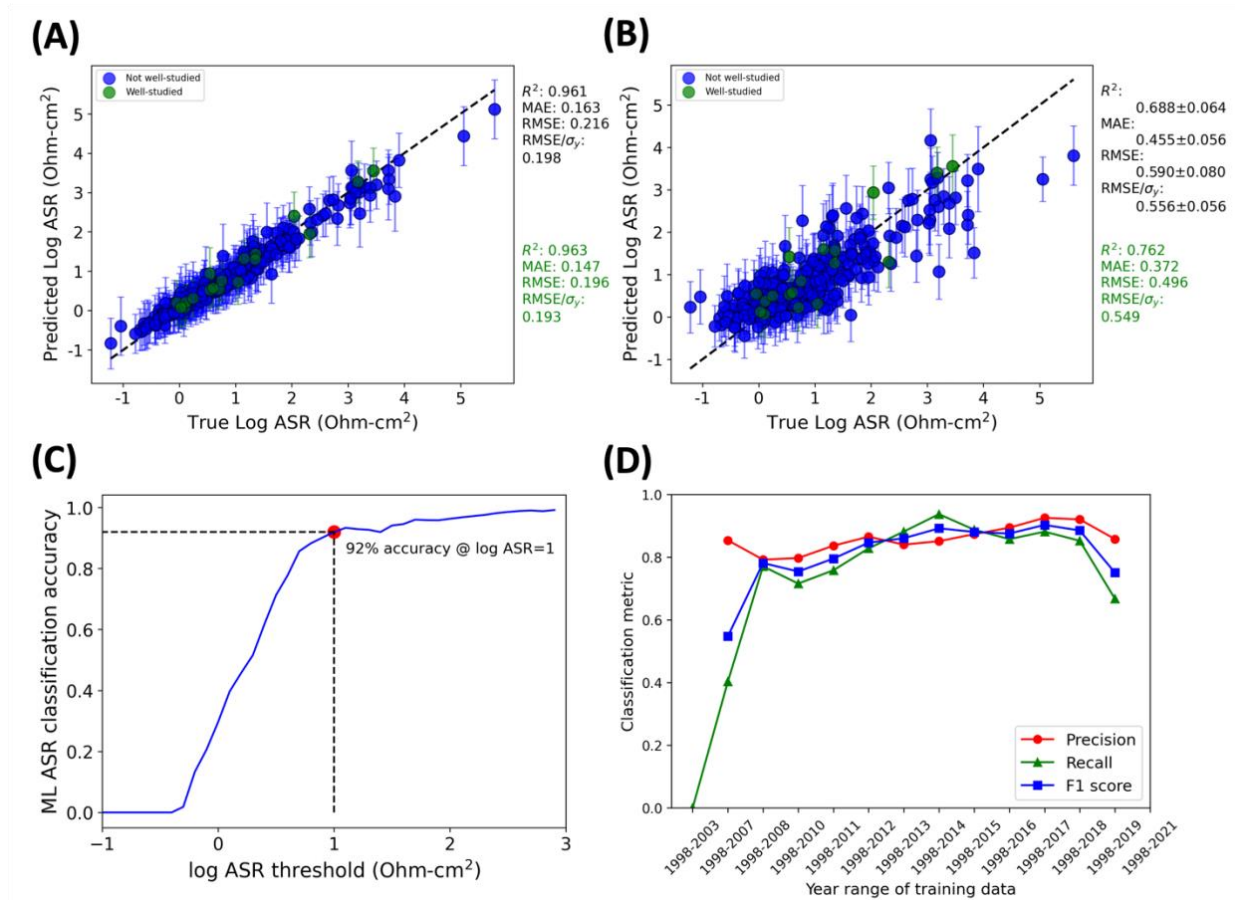

**Figure 2.** Summary of ML model performance for predicting log ASR at T = 500 °C. (A) Parity plot of full-fit to all of the data, (B) 5-fold CV assessment, (C) ASR model classification accuracy for predicting materials with log ASR < 1 Ohm-cm$^2$, (D) temporal cross validation classification assessment. In (A) and (B), the error bars are the recalibrated random forest ensemble error bars. In (A) and (B), the metrics listed in black are assessments on all of the data, where the +/- values are the standard deviation over all CV splits, and the metrics listed in green are assessments on the subset of 18 well-studied materials with greater than 4 experimental measurements.

Here, we compare the performance of our ML model for predicting ASR with that from recent work by Zhai et al.,[34] who used a database of 85 ASR values and a neural network ML model employing similar elemental features as used here, together with a newly-proposed ionic Lewis acid descriptor, to screen for new perovskite materials based on low predicted ASR values. Zhai et al. obtained a log ASR RMSE of 0.336 Ohm-cm$^2$ for data at 700 °C. We note that this RMSE of 0.336 Ohm-cm$^2$ is lower than our current random forest model on our ASR database, which yielded an RMSE of 0.590 +/- 0.080 Ohm-cm$^2$ for data at 500 °C. However, this comparison uses different ASR data, a different model, and a different feature set, so this comparison cannot be used to assess the relative merits of the two approaches. To allow a more direct comparison, we use a random forest model and our feature generation scheme (e.g., the ionic Lewis acid descriptor used in the work of Zhai et al. is not included) and perform 5-fold CV on the same ASR dataset used in the work of Zhai et al. From this analysis, we obtain an RMSE of 0.367 +/- 0.049 Ohm-cm$^2$. The RMSE of 0.336 Ohm-cm$^2$ quoted in Zhai et al. is within the CV sampling error bar of our present result, suggesting that the main effect of the different RMSE values between their study and the present work is likely the ASR dataset used for train/test as opposed to the model type or feature set used. It is helpful to compare the reduced RMSE/σ (σ = dataset standard deviation) instead of just RMSE values, in order to normalize the predictive performance by the spread of data used in the model. The standard deviation of the log ASR values used by Zhai et al. is 0.505 Ohm-cm$^2$, producing an RMSE/σ value of 0.665. The log ASR database used for the fit in **Figure 2** has a standard deviation of 1.083 Ohm-cm$^2$, producing, from 5-fold CV, an RMSE/σ = 0.556 +/- 0.056. By this measure our model has a modestly lower average reduced RMSE than Zhai et al.'s model.

## Insights into role of featurization and feature importances of the ASR ML model

The work of Zhai et al. and the present work highlight two different featurization approaches for composition-based materials property prediction. The first approach involves the use of human effort and physical intuition to craft physics-based features like the Lewis acid descriptor or O p-band center. The second approach is more data-centric and involves providing a full suite of trivial-to-calculate elemental features and letting the ML model select the most important features. There are pros and cons to both approaches. The first approach utilizes physical understanding of how the features relate to the target property and therefore may be better able to represent target data with fewer features than the data-centric methods. This property of physics-based features is likely to be particularly advantageous for smaller datasets, where there may not be enough data for the ML model to select the most effective features in the more data-centric approaches. Furthermore, physics-based features often provide physical insight, e.g., by examining how the target depends on the feature. However, physics-based features are often harder to develop, as they require significant domain expertise, and often harder to implement, as they can require more work to calculate for new test data (e.g., the Lewis acid descriptor described above requires knowledge of the oxidation state of each element in the material, which must be estimated based on additional chemical models). In contrast, the data-centric approach uses a large suite of tabulated elemental features that can be developed with no domain knowledge, are easy to generate, and are easy to implement on new test data. The large number of features may require a relatively larger dataset than the physics-based approaches to extract the important features and their relationships to the target. However, a large set of simple elemental features may be advantageous compared to physics-based features on large data sets as they provide enormous flexibility and do not presuppose any particular physical mechanism. Even if the elemental features produce an excellent ML model, it may be difficult to establish physical understanding of why the features influence the target property in certain ways or develop useful mechanistic understanding from the ML model. Furthermore, having more features will slow down model fits, which may present an issue for some use cases. The present work and work of Zhai et al. do not use purely physics-based or data-centric features, but instead leverage some combination of the two. The work of Zhai et al. relies predominantly on the use of physics-based features, though they also include some additional elemental features motivated by physical intuition. The present work relies predominantly on the use of data-centric elemental features, though we also include the one-hot electrolyte encoding and ML-predicted ASR barrier as a feature (note the ASR barrier model uses elemental features and one-hot electrolyte encoding as features), which means we also have some physics-based features in our work.

We tested the addition of the Lewis acid descriptor in our present approach and found that our random forest model did not find this feature to significantly impact the results (i.e., it had a low feature importance), thus the model performance was unchanged. We attribute the low feature importance of the Lewis acid descriptor in our case to be the result of the large set of elemental features available for the model, which effectively contribute the same information as the Lewis acid descriptor. Therefore, this result does not imply that the Lewis acid descriptor does not contribute important physical information. In fact, we performed a test using a feature set consisting of only the ML-predicted ASR barrier, A- and B-site Lewis acid descriptors, ML-predicted O p-band center, and one-hot electrolyte encoding, for a total of just 8 features. We find a 5-fold CV MAE, RMSE, and RMSE/σ of 0.463 +/- 0.047 Ohm-cm$^2$, 0.610 +/- 0.061 Ohm-cm$^2$, and 0.575 +/- 0.079, respectively (the +/- is standard deviation over 25 splits). These 5-fold CV average values are slightly higher than our best model shown in **Figure 2**, but are within the CV sampling standard deviation and demonstrates the power of only a handful of physically-motivated features to create a simple and quite accurate model. It is worth noting that if the A- and B-site Lewis acid descriptors are removed, the 5-fold CV model error increases modestly, where the MAE, RMSE, and RMSE/σ of 0.509 +/- 0.042 Ohm-cm$^2$, 0.656 +/- 0.062 Ohm-cm$^2$, and 0.615 +/- 0.057, respectively (the +/- is standard deviation over 25 splits), indicating the addition of the Lewis acid descriptor is helpful for forming an accurate ASR model if one is interested in using a small, physically-motivated feature set. Overall, these results suggest that the simple elemental features used here are as good or better than the Lewis acid feature from Zhai et al. for prediction of ASR, although the Lewis acid feature represents a much more physically intuitive and economical featurization.

To extract some understanding of how the features explored in this study relate to the target and develop more physical intuition from the ML models, we here examine the relative importance of the input features in predicting ASR using the SHapley Additive exPlanation (SHAP) approach[42] with a random forest model, as shown in **Figure 3.** Briefly, SHAP is a mathematical method of explainable machine learning which uses game theoretic principles to assess the contributions of each feature in a model prediction, thus providing both ranked feature importances and trends of the target variable with each feature.

In **Figure 3A,** we examine the SHAP ranking for the model discussed in the preceding paragraph, which integrated the Lewis acid descriptor with elemental feature-based models of ASR energy barrier and O p-band center, and one-hot electrolyte encoding. There are clear physical trends of these features with the resulting ASR values. For example, higher values of activation energy tend to coincide with materials with higher ASR values. This trend makes intuitive sense, because for ASR the Arrhenius scaling relationship is ASR $\propto$ exp($E_a$/kT), where $E_a$ is the activation barrier, $k$ is the Boltzmann constant and $T$ is the absolute temperature, where larger $E_a$ produces a larger ASR, because ASR increases with decreasing $T$. In addition, a lower (higher) A-site (B-site) Lewis acid value is indicative of a material with weaker metal-oxygen bonding, promoting lower ASR. For example, $La^{3+}$ and $Ba^{2+}$ have Lewis acid values of 0.343 and 0.194, respectively, and broadly, A-site Ba-based perovskites have lower ASR than A-site La-based perovskites). As another example for the B-site, $Mn^{3+}$ and $Co^{4+}$ have Lewis acid values of 0.513 and 0.666, respectively, and Co-based perovskites generally have lower ASR than Mn-based perovskites. It is worth noting that recent work from Xu et al. used knowledge of low Lewis acid strength $Cs^+$ cation on the A-site to suggest $PrBa_{0.9}Cs_{0.1}Co_2O_{5+\delta}$ is a promising PCFC electrode material.[43] In addition, higher values of O p-band center correspond to lower ASR values, again consistent with the fact that higher O p-band center corresponds to weaker oxygen binding in a perovskite material. Finally, the electrolyte encoding offers some useful insights as well. For example, materials with a zirconia or perovskite electrolyte tend to correspond to higher ASR values, while materials with a ceria electrolyte tend to correspond to lower ASR values.

In **Figure 3B,** we examine the SHAP ranking (showing the top 25 of 50 features) for our best model discussed in the context of **Figure 2**. Similar to the other model, the ASR activation barrier is the most important feature. The trends with electrolyte type are also the same between the two models. For this model, nearly all of the top elemental features (denoted here as "*{property_name}_{math_operation}*") generally fall into the following groups of properties: features related to size and/or volume (e.g., GSvolume_pa_composition_average, AtomicVolume_arithmetic_average, ICSDVolume_composition_average, etc.), features related to thermodynamics of formation and thus bond strength (e.g., HeatFusion_composition_average, BoilingT_arithmetic_average, etc.), features related to the electronic structure or electronic properties of constituent elements (e.g., valence_composition_average, phi_arithmetic_average, phi_difference, etc.), or features which help to distinguish different chemistries of the periodic table (e.g., IsRareEarth_composition_average, IsFBlock_composition_average). The ML model selection of many features comprising an array of different physical and chemical properties is reflective of our above discussion of the trade-offs between physics-based and data-centric approaches of feature generation, where the present approach using many elemental features can well-represent complex physical relationships, but at some cost of interpretability. While we don't seek to have a deep understanding of the physical trends of each elemental feature selected by our model, some useful trends can still be extracted. For example, the higher values of the features IsRareEarth_composition_average, IsFBlock_composition_average corresponding to higher ASR values makes sense, as those perovskites that contain mostly rare earth elements on the A-site tend to have higher ASR values than those that are, e.g., mostly alkaline earth elements. Higher values of GSvolume_pa_composition_average (GS = ground state volume), ICSDVolume_composition_average, and CovalentRadii_composition_average generally correspond to lower ASR values, which we speculate may be the result of including large alkaline earth elements (e.g., Sr, Ba) at the expense of smaller rare earth elements (e.g., La, Y, Pr), producing a larger volume and more cubic perovskite lattices. While the trend of larger AtomicVolume_arithmetic_average corresponding to larger ASR may appear an exception to this, we observe this trend is consistent with the fact that moving from early to late transition metals (e.g., Ti to Co) results in smaller transition metal atomic volume, which is consistent with lower ASR values. Finally, the larger values of phi_arithmetic_average and phi_difference (note, here "phi" means work function) being indicative of lower ASR values makes sense again through the lens of O binding strength and our previous studies on correlating work function with O p-band center. Namely, materials with high work functions are easier to electrochemically reduce than those with low work function, where the ease of reduction implies weaker metal-oxygen bonding, hence resulting in lower ASR values. Consistent with the above discussion, it is difficult to extract physical meaning from the trends with basic elemental features. However, as noted above, there simplicity and accuracy make them a very appealing option.

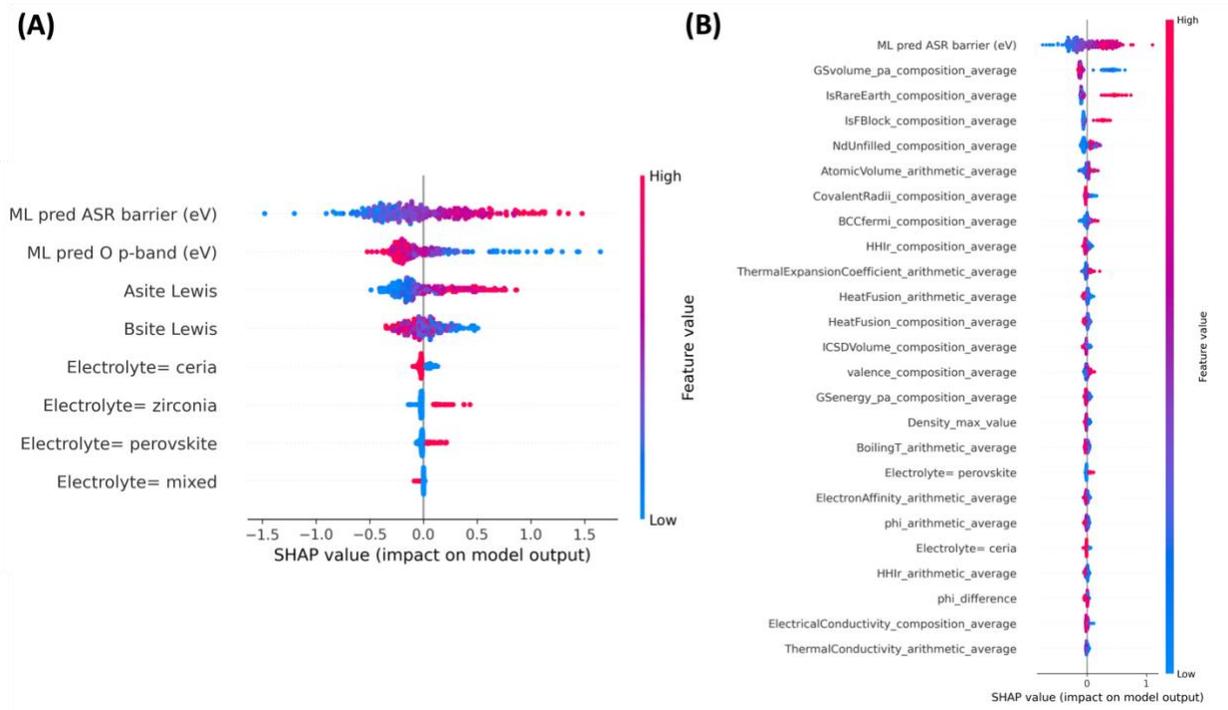

**Figure 3.** Summary of feature importances for our ASR model using the SHAP method. (A) SHAP feature ranking for ASR model constructed using physically-motivated features. (B) SHAP feature ranking of the top 25 features of our best performing elemental feature-based model from **Figure 2**.

## Screening new promising perovskite catalysts with machine learning

In this section, we use our ML model for ASR discussed above, together with calculations of material cost (calculated using the pymatgen[44] package) and separate ML model predictions of perovskite stability to screen for new promising perovskite catalysts. The stability model is made using a random forest model and elemental features to predict values of 2844 perovskite oxides measured as convex hull energy using the database from Ma et al.[21] (who expanded and updated the database from Li et al.[45], more details are given in **Section S4** of the **SI**). In total, we enumerate a large search space consisting of up to 3 elements on the A-site and 4 elements on the B-site, covering 50 elements and totaling just over 19M (19,072,821) materials (see **Section S5** of the **SI** for more details). To find new promising materials, we set our screening criteria to be values below a target threshold of cost, stability, and ASR activity. We set the threshold value of cost to coincide with the value of the commercial material $La_{0.6}Sr_{0.4}Co_{0.2}Fe_{0.8}O_3$ (LSCF), which is $133.67 per kg. For activity, we set the threshold of the log ASR value of LSCF (at 500 °C) of 1.33 Ohm-cm$^2$, and the log ASR (at 500 °C) of representative top performing material Ba$_{0.5}$Sr$_{0.5}$Co$_{0.8}$Fe$_{0.2}$O$_3$ (BSCF), which is 0.21 Ohm-cm$^2$. The threshold for stability is informed by the work of Zhai et al. From their study, their top-performing material, Sr$_{0.9}$Cs$_{0.1}$Co$_{0.9}$Nb$_{0.1}$O$_3$ (SCCN), was demonstrated to have stable operation at 550 °C for over 800 hours without any observed loss in performance. Our stability model predicts a value of 93.3 meV/atom for SCCN at 500 °C, which we use as the stability threshold for screening. The order under which each property is screened can impact the resulting lists at different stages of the screening (although the resulting list when all criteria are applied does not depend on the order of their application) and multiple orders are explored below. Throughout this analysis, we make predictions of ASR assuming a ceria electrolyte is used.

**Figure 4** contains violin plots showing the distributions of cost (**Figure 4A**), stability (**Figure 4B**), and predicted log ASR at 500 °C (**Figure 4C**) as the screening criteria are successively applied, starting with the criteria being plotted in each case. From **Figure 4**, we can see that 2,453,872, 1,393,424 and 2,135,396 materials separately pass the screening criteria of cost, stability, and ASR, respectively, which translates into 12.9%, 7.3% and 11.2% of the original 19,072,821 considered materials, with stability being the most stringent screening criterion. A total of 57,579 materials (0.30%) pass both the cost and stability screening, while 53,210 (0.28%) materials pass both the stability and ASR screening. Finally, 9135 (0.05%) materials pass all screening criteria. A spreadsheet containing the compositions, calculated costs, and predicted stability and ASR values for these 9135 materials is provided as part of the **SI**.

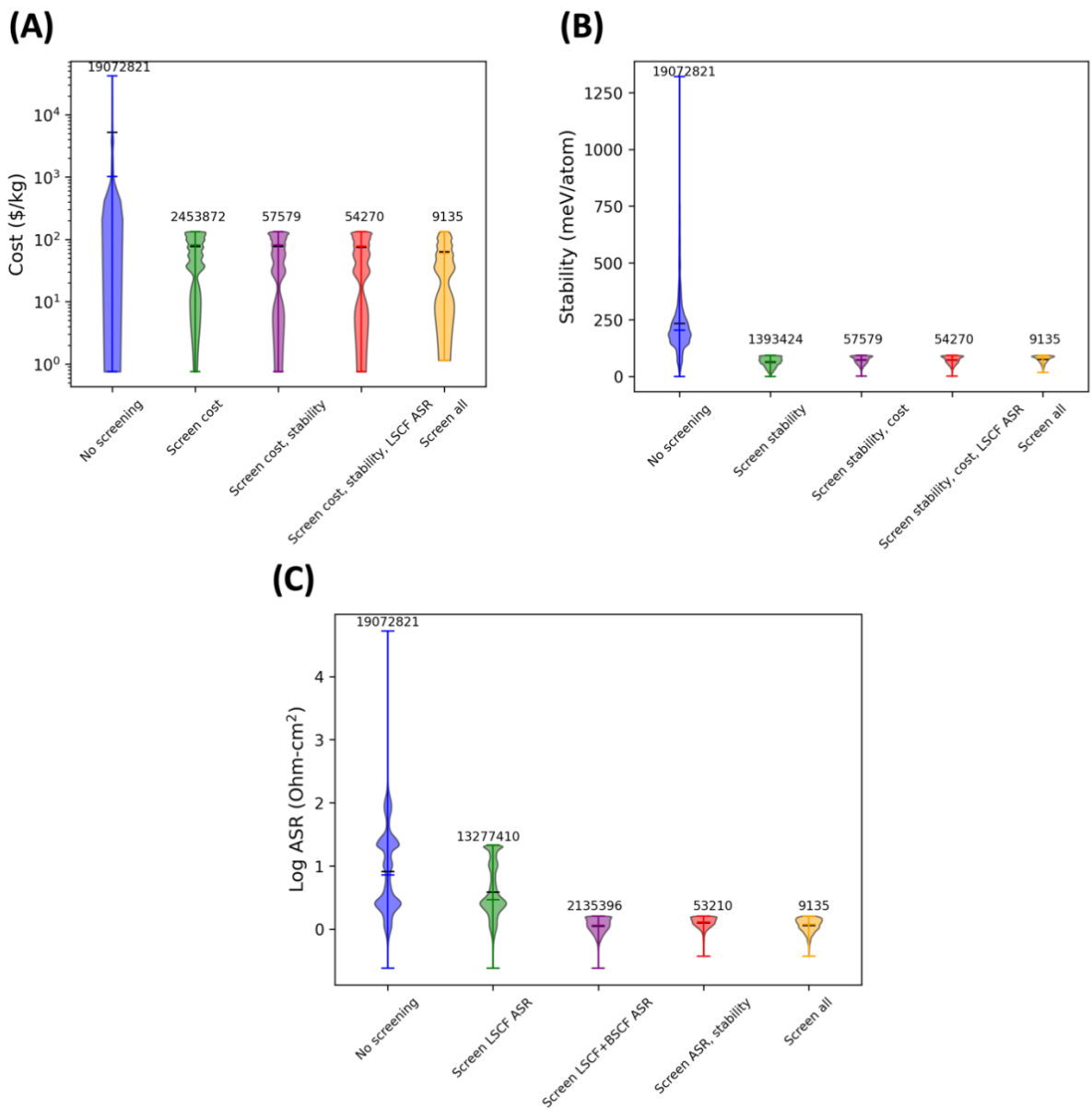

**Figure 4.** Violin plots showing distributions of screened materials where the first screening is (A) screened materials cost, (B) screened materials stability, and (C) screened ASR. The numbers above each distribution denote the number of materials passing the given screening combination. The high, middle, and low colored ticks denote the maximum, median and minimum of the distribution, respectively, while the black ticks denote the mean of the distribution.

We can use our ASR model and our list of predictions to make numerous material assessments and comparisons. First, we can examine the most favorable screened materials from each of our screening criteria of cost, stability and ASR value. From inspecting this list of screened promising materials, we find that the materials that are the cheapest, most stable, and most highly active are BaFe$_{0.75}$Cu$_{0.125}$Zr$_{0.125}$O$_3$ ($1.15/kg, log ASR at 500 °C = 0.12 Ohm-cm$^2$), BaFe$_{0.5}$Co$_{0.25}$Mo$_{0.25}$O$_3$ (18.0 meV/atom, log ASR at 500 °C = -0.02 Ohm-cm$^2$) and SrCo$_{0.75}$Nb$_{0.125}$Ta$_{0.125}$O$_3$ (SCNT) (log ASR at 500 °C = -0.43 Ohm-cm$^2$), respectively. It is worth noting that the most active material from our screening, SCNT, is already known to be high performing. Second, we can examine our list of low ASR materials to search for new materials that are compositionally distinct from known materials, making them interesting for further study. Inspecting the top candidates in our list for materials which are high performing and compositionally unique compared to known materials reveals materials with unusual combinations of elements like K, Bi, Y, Ni and Cu, indicating relatively unexplored compositions that may be worth further attention. For example, the materials SrZr$_{0.125}$Nb$_{0.125}$Co$_{0.625}$Cu$_{0.125}$O$_3$ (SZNCCu), K$_{0.25}$Sm$_{0.125}$Sr$_{0.625}$Nb$_{0.125}$Ta$_{0.125}$Co$_{0.75}$O$_3$ (KSmSCNT) and Bi$_{0.125}$Sr$_{0.875}$Y$_{0.125}$Ni$_{0.125}$Co$_{0.75}$O$_3$ (BiSYNC) have very low predicted log ASR values at 500 °C of just -0.37, -0.33 and -0.25 Ohm-cm$^2$, respectively. Third, we can use our model to evaluate other recently proposed promising materials from the work of Zhai et al. and compare them with the materials proposed here. Our model predicts all four promising materials from the work of Zhai et al. to have log ASR values lower than BSCF at 500 °C, consistent with experimental validation by Zhai et al. finding all four of these materials are indeed more active than BSCF. These 4 materials are SCCN, Ba$_{0.4}$Sr$_{0.4}$Cs$_{0.2}$Co$_{0.6}$Fe$_{0.3}$Mo$_{0.1}$O$_3$ (BSCCFM), Ba$_{0.8}$Sr$_{0.2}$Co$_{0.6}$Fe$_{0.2}$Nb$_{0.2}$O$_3$ (BSCFN), and Sr$_{0.6}$Ba$_{0.2}$Pr$_{0.2}$Co$_{0.6}$Fe$_{0.3}$Nb$_{0.1}$O$_3$ (SBPCFN). Our approach indicates the cost, stability, and predicted log ASR of these materials are SCCN: ($4337.37/kg, 93.2 meV/atom, 0.02 Ohm-cm$^2$), BSCCFM: ($7749.79/kg, 137.1 meV/atom, -0.09 Ohm-cm$^2$), BSCFN: ($7.90/kg, 92.0 meV/atom, -0.03 Ohm-cm$^2$), SBPCFN: ($77.75/kg, 130.1 meV/atom, -0.06 Ohm-cm$^2$).

Since our ASR model uses the ML-predicted activation energy as a feature, we can predict ASR as a function of temperature, as shown in **Figure 5**. **Figure 5** shows predicted ASR values (solid lines) with calibrated error bars as a function of temperature together with experimental data points for the commercial material LSCF, benchmark high performing material BSCF, the best material from Zhai et al., SCCN, and the top selected materials from our screening: SZNCCu, KSmSCNT, and BiSYNC. In the low ASR regime, our ASR model has a tendency to predict higher than true ASR values (a slight conservative bias), where this bias is about 0.3 log units for LSCF, BSCF and SCCN averaged together. However, the data points are within the calibrated uncertainties of the ML model. Our screened materials SZNCCu, KSmSCNT, and BiSYNC are predicted to outperform BSCF and SCCN at 500 °C. In addition, they have lower activation barriers than BSCF, implying their performance will continue to outpace BSCF for temperatures below 500 °C, shown in **Figure 5** via extrapolation to 400 °C.

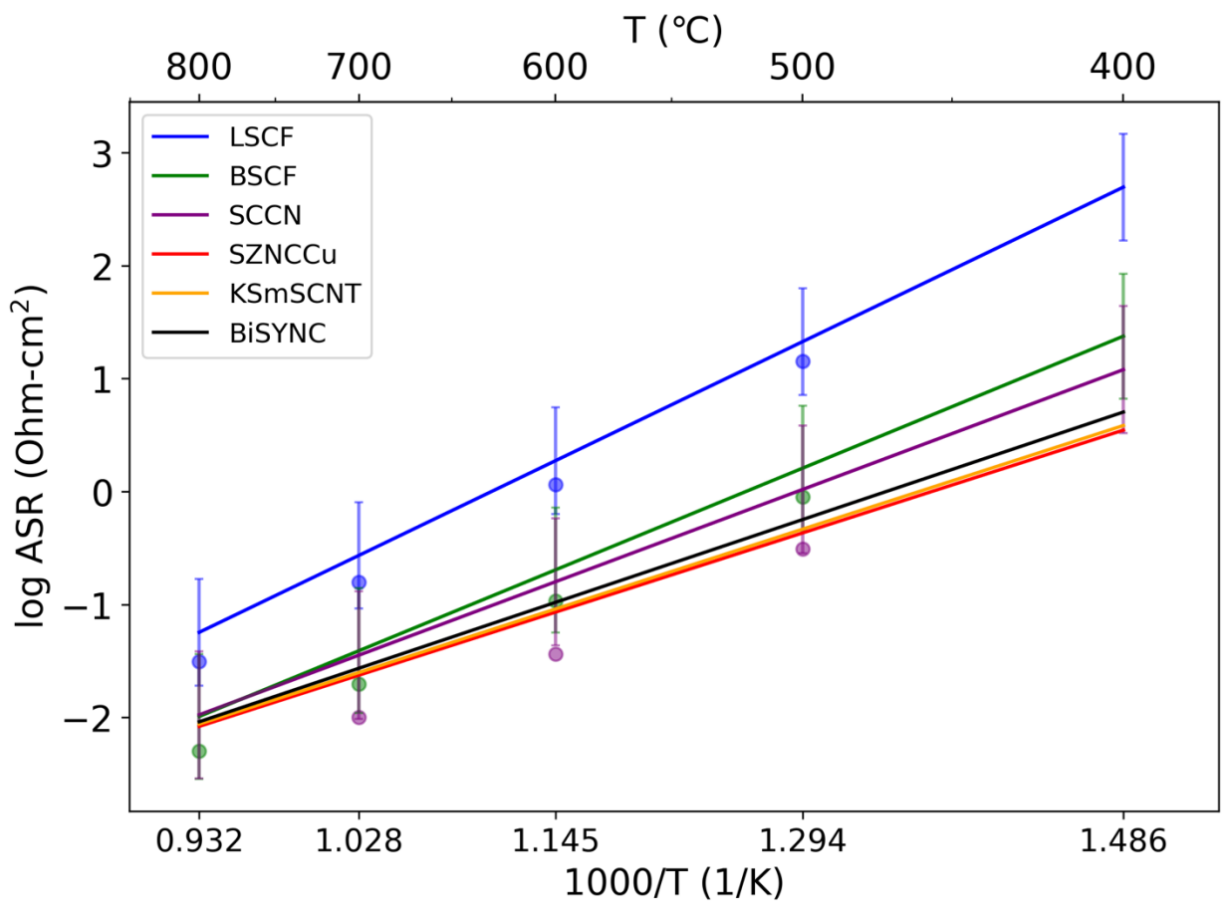

**Figure 5.** ML-predicted ASR temperature dependence for key materials. The solid lines are ML predictions using our predicted log ASR value at 500 °C together with the ML model of predicted ASR barrier to scale the prediction to other temperatures. The error bars are the calibrated one standard deviation error bars from the ML model. Data points are experimental ASR values extracted from the database from Jacobs et al.[39] (LSCF, BSCF) and from Zhai et al.[34] (SCCN).

## Summary and conclusions

In this work, we developed a fully data-centric ML approach to predicting perovskite oxygen catalytic and transport properties by leveraging the largest (at the time of this writing) database of perovskite oxygen catalytic properties comprising oxygen surface exchange rates ($k_{chem}$ and $k^*$), oxygen diffusivity ($D_{chem}$ and $D^*$), and ASR data. We show that random forest ML models fit using trivial-to-obtain elemental features can produce cross validation average MAE values either on par or lower than those obtained using linear correlations of the DFT-calculated O p-band center. These ML models are orders of magnitude faster to evaluate than using the O p-band center descriptor as a DFT calculation for each material is not needed, providing a path toward fast screening of perovskite catalytic properties. Using our ML model of ASR, we screen a set of over 19M perovskite compositions and propose numerous new promising materials that are cheaper than the commercial material LSCF, more stable than the well-performing material SCCN, and predicted to have exceptional low ASR values at T = 500 °C, making them worthy of additional study. Finally, our ASR model is equipped with well-calibrated uncertainty estimates to foster further confidence in the property predictions, and is accessible online for broader use by the community.

**Acknowledgements:** This project was funded by the United States Department of Energy, National Energy Technology Laboratory, in part, through a site support contract. Neither the United States Government nor any agency thereof, nor any of their employees, nor the support contractor, nor any of their employees, makes any warranty, express or implied, or assumes any legal liability or responsibility for the accuracy, completeness, or usefulness of any information, apparatus, product, or process disclosed, or represents that its use would not infringe privately owned rights.  Reference herein to any specific commercial product, process, or service by trade name, trademark, manufacturer, or otherwise does not necessarily constitute or imply its endorsement, recommendation, or favoring by the United States Government or any agency thereof. The views and opinions of authors expressed herein do not necessarily state or reflect those of the United States Government or any agency thereof.

**Conflicts of Interest**

The authors of have no conflicts of interest to declare.

**Data Availability**

The spreadsheet of all calculated cost, predicted stability and activity values for the full 19M materials prediction dataset, as well as the screened set of top materials, is available on Figshare: https://doi.org/10.6084/m9.figshare.24450445.v1.[46] The final ASR model is available on Github: https://github.com/uw-cmg/ASR_model. The model can be run on Google Colab via the link in the Github repository. This model provides predicted ASR values at 500 °C with calibrated uncertainties, and predicted stability values and calculated costs. Since the model also produces a predicted activation energy, the predicted ASR values at 500 °C can be scaled to the desired temperature of interest.

# Supplementary Information for

**Machine Learning Design of Perovskite Catalytic Properties**

**Authors:** Ryan Jacobs[1,*], Jian Liu[2,*], Harry Abernathy[2], Dane Morgan[1]

[1] Department of Materials Science and Engineering, University of Wisconsin-Madison, Madison, WI, 53706, USA.

[2] National Energy Technology Lab, Morgantown, WV, 26505, USA.

# Section S1: Additional details of machine learning model development and assessment

All machine learning model fits and assessments in this work were performed using the MAterials Simulation Toolkit for Machine Learning (MAST-ML).[40] Briefly, MAST-ML simplifies and accelerates the use of machine learning models in materials science, enabling users to quickly perform assessments of multiple models and cross validation tests, with codified analysis data and figure output. MAST-ML leverages widely used machine learning libraries like scikit-learn[47] and Keras[48] (Tensorflow[49] backend). For the catalytic property fits shown in **Figure 1** of the main text, we evaluated the efficacy of random forests, gradient boosting regressors, and extreme gradient boosting (XGBoost) models in predicting the catalytic property values. We generally found these three models showed comparable performance, with random forest showing slightly better performance across all properties. The perovskite compositions were featurized using elemental properties from the *ElementalFeatureGenerator* class in MAST-ML. Briefly, this feature generator uses a set of elemental properties originally from the MAGPIE database[50] but expanded in works by Wu et al.[51] and Lu et al.,[52] and includes basic arithmetic operations on those features such as maximum, minimum, difference, and composition-weighted average. For the case of machine learning models of ASR, the electrolyte used in the measurement (i.e., zirconia-based for YSZ and ScSZ, ceria-based for GDC and SDC, perovskite for LSGM) was also included as a feature using one-hot encoding. The most pertinent set of features were selected using the ranked feature importances from a random forest model.

For the fits to our full ASR database in **Figure 2** of the main text, we focused on using random forest models and evaluated the 5-fold CV fit quality of different feature sets. We compared the performance using (i) only elemental features, (ii) elemental features plus one-hot encoding of electrolyte type (as was done in the analysis of **Figure 1** of the main text), (iii) elemental features, one-hot encoding of electrolyte type, and the ML-predicted ASR activation barrier. The ASR barrier was predicted using a separate random forest model trained using elemental features plus one-hot encoding. We found that adding the one-hot encoding of electrolyte and the ML-predicted ASR barrier had about a 10% improvement in average 5-fold CV metrics, as summarized in **Table S1**.

**Table S1.** Summary of random forest ML model 5-fold CV performance for predicting log ASR at 500 °C using different feature sets. The quoted values are averages from 5-fold CV +/- standard deviation over 25 folds.

| Feature set | 5-fold CV MAE | 5-fold CV RMSE | 5-fold CV RMSE/σ |
|---|---|---|---|
| (i) | 0.509 +/- 0.050 | 0.662 +/- 0.080 | 0.624 +/- 0.061 |
| (ii) | 0.491 +/- 0.058 | 0.640 +/- 0.078 | 0.601 +/- 0.063 |
| (iii) | 0.454 +/- 0.047 | 0.596 +/- 0.074 | 0.564 +/- 0.066 |

## Section S2: Additional details of machine learning analysis

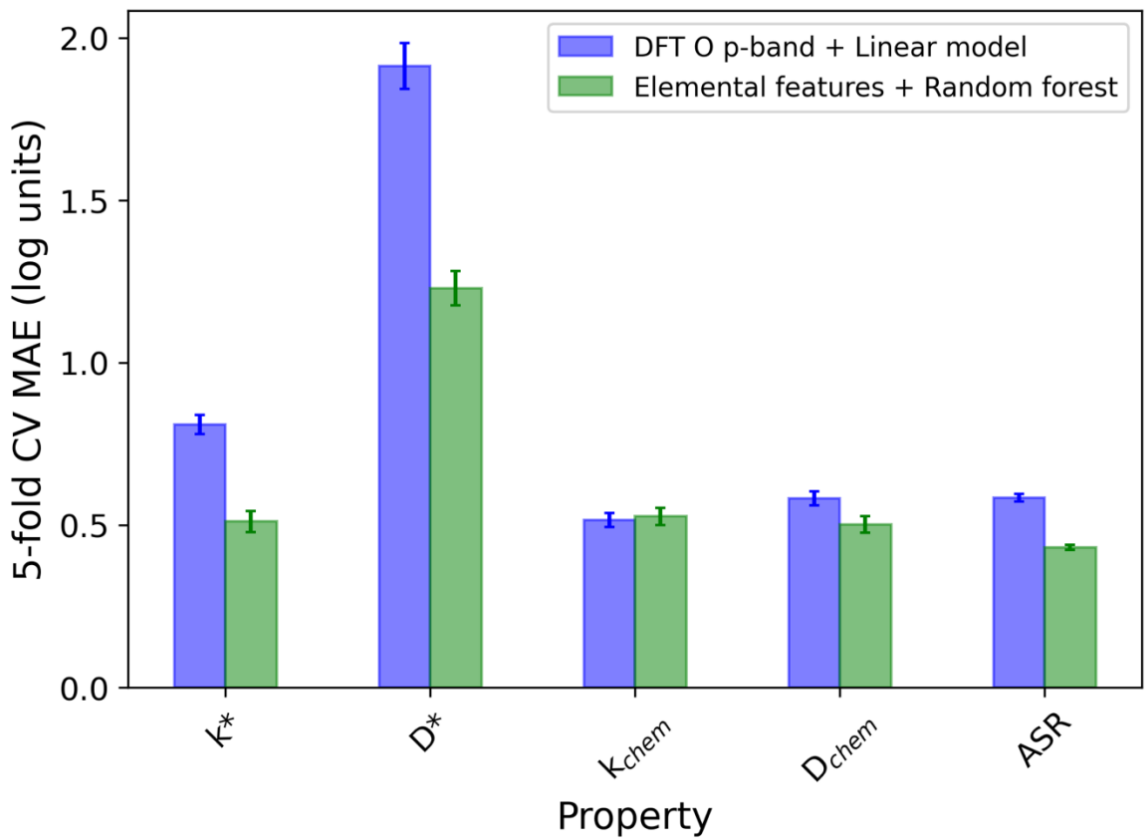

**Figure S6.** Machine learning model random cross validation assessment comparing performance of the DFT-calculated O p-band center descriptor with a linear model and a random forest model using elemental features. These assessments are for T = 800 °C. The units of k* and $k_{chem}$ are cm/s, the units of D* and $D_{chem}$ are $cm^2/s$, and the units of ASR are Ohm-$cm^2$. The error bars are standard errors in the mean of the calculated MAE over 25 splits of 5-fold CV.

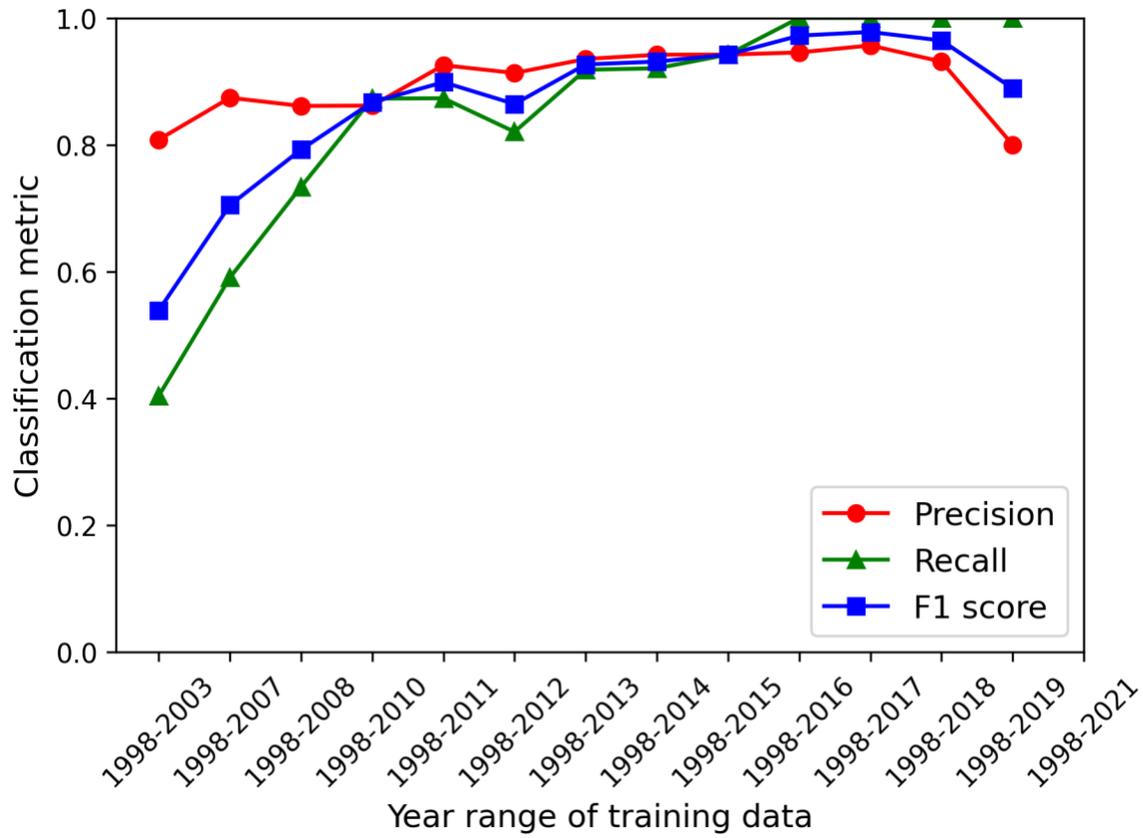

**Figure S7.** Temporal cross validation results for ASR model fit at 800 °C.

## Section S3: Additional details of error bar analysis and recalibration

**Figure S8** contains an assessment of the quality of the uncertainty estimates (error bars) of our ML model using what we call a "residual vs. error" (RvE) plot.[26,41] In order for the uncertainty estimates (i.e., predicted errors) to be accurate, the distribution of their values should match that of normalized residuals (i.e., actual errors), meaning the slope of the fit line in **Figure S8** should be one and the intercept should be zero. In **Figure S8**, we can see that the uncalibrated uncertainty estimates (grey points and fit line) significantly underestimate the true error for small residuals, and significantly overestimate the error for large residuals, resulting in a low slope of 0.50 and high intercept of 0.28. The data after calibration (blue points and fit line) show significant improvement, with a slope of 0.98 and intercept of 0.01. This result shows the error bars are more accurate after recalibration, though they are not perfect, and may be improved as additional data becomes available or new uncertainty estimate approaches are used.

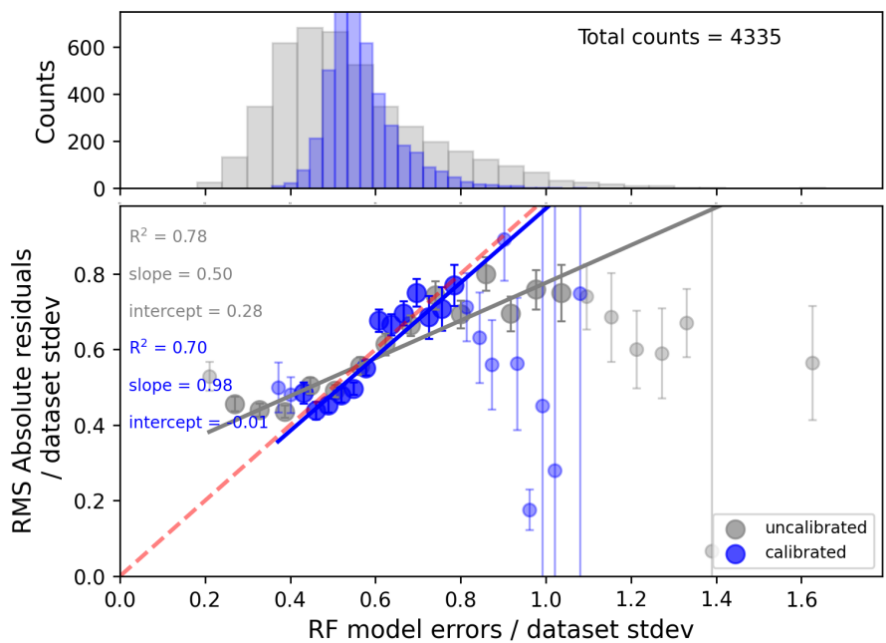

**Figure S8.** Summary of the random forest uncertainty estimates and their recalibration. This model is for predicting ASR at 500 °C and uses elemental features, one-hot electrolyte encoding and ML-predicted ASR barrier as features. The average (+/- standard deviation) recalibration parameters are $a$ = 0.42824 +/- 0.10199 and $b$ = 0.36342 +/- 0.05895.

## S4. Additional details of stability database

To formulate a model for perovskite stability, we use the dataset of perovskite oxides from the work of Ma et al.[21] This database contains DFT-calculated total energies of 2935 perovskites. After removing duplicate entries (the duplicate entry with higher DFT total energy was removed) and those entries with very high DFT energies (indicating the run did not converge), the number of entries amounts to 2844 materials. The stability (as energy above the convex hull) of each perovskite was calculated using pymatgen.[44] More specifically, we calculate the stability in both a closed system (using the PhaseDiagram() class in the pymatgen.analysis.phase_diagram module) and a system open to O (using the GrandPotentialPhaseDiagram() class in the pymatgen.analysis.phase_diagram module). The stability values for the system open to O were used for the stability ML model for screening new materials in the main text. For the stability calculations, the materials were queried from the Materials Project using the latest version of the API through the mp-api package. To account for the mixing of GGA and GGA+$U$ energies, the energy corrections updated in late 2020 were used, and are tabulated in the MP2020Compatibility.yaml file in the pymatgen.entries folder of the pymatgen Github page. These corrections consist of: V = -1.7 eV/V, Cr = -1.999 eV/Cr, Mn = -1.668 eV/Mn, Fe = -2.256 eV/Fe, Co = -1.638 eV/Co, Ni = -2.541 eV/Ni, W = -4.438 eV/W, Mo = -3.202 eV/Mo, O = -0.687 eV/O. The O chemical potential was calculated at a temperature of 500 °C and pressure of 0.2 atm, and included basic vibrational corrections following previous work.[53,54] The value used for the O chemical potential was -5.774 eV/O. The energies above the convex hull were normalized to the 40-atom perovskite formula units used in the work of Ma et al.[21]

## S5. Additional details of materials screening

We enumerated a large prediction space of roughly 19 million distinct perovskite compositions covering 50 elements. Perovskite A-site elements were chosen from the set of alkali, alkaline earth and rare earth elements, while perovskite B-site elements were chosen from the set of transition metal, post-transition metal and redox inactive elements. Perovskite compositions were constructed based on different subsets, then combined. The first set consisted of $ABO_3$, $AA'BO_3$, $ABB'O_3$ and $AA'BB'O_3$ compositions where A-site was chosen from {Mg, Cs, Sr, Ba, Cd, Zn, Bi, La, Ce, Pr, Nd, Sm, Eu, Gd, Dy, Ho, Yb, Lu} and B-site chosen from {Sc, Y, Al, Ga, In, Ti, V, Cr, Mn, Fe, Co, Ni, Cu, Si, Ge, Sn, Pb, Zr, Nb, Mo, Ru, Rh, Pd, Hf, Ta, W, Re, Os, Ir, Pt} and the composition increments were x = y = 0.125, and sum(A-site) = sum(B-site) = 1. This initial set resulted in 3,348,675 compositions. The remaining sets focused on the subset of elements where A-site was chosen from {K, Rb, Cs, Ca, Sr, Ba, Bi, La, Ce, Pr, Nd, Sm} and B-site was chosen from {Sc, Y, Ce, Ti, Fe, Co, Ni, Cu, Sn, Zr, Nb, Mo, Ta}. The second set consisted of $ABB'B''O_3$ materials and resulted in 78,780 compositions. The third set consisted of $AA'A''BB'B''O_3$ materials and resulted in 15,356,796 compositions. The final set consisted of $ABB'B''B'''O_3$ materials and resulted in 378,924 compositions. After removing duplicates obtained after combining the various subsets, the final list consists of 19,072,821 compositions.

# References

(1) Ndubuisi, A.; Abouali, S.; Singh, K.; Thangadurai, V. Recent Advances, Practical Challenges, and Perspectives of Intermediate Temperature Solid Oxide Fuel Cell Cathodes. *Journal of Materials Chemistry A*. Royal Society of Chemistry February 7, 2022, pp 2196–2227. https://doi.org/10.1039/d1ta08475e.

(2) Jacobson, A. J. Materials for Solid Oxide Fuel Cells. *Chemistry of Materials* **2010**, *22*, 660–674. https://doi.org/10.1021/cm902640j.

(3) Haile, S. M. Fuel Cell Materials and Components. *Acta Mater* **2003**, *51*, 5981–6000. https://doi.org/10.1016/j.actamat.2003.08.004.

(4) Wachsman, E. D.; Lee, K. T. Lowering the Temperature of Solid Oxide Fuel Cells. *Science (1979)* **2011**, *334* (6058), 935–939. https://doi.org/10.1126/science.1204090.

(5) Duan, C.; Tong, J.; Shang, M.; Nikodemski, S.; Sanders, M.; Ricote, S.; Almansoori, A.; O'Hayre, R. Readily Processed Protonic Ceramic Fuel Cells with High Performance at Low Temperatures. *Science (1979)* **2015**, *349* (6254), 1321–1326. https://doi.org/10.1126/science.aab3987.

(6) Li, Z.; Li, M.; Zhu, Z. Perovskite Cathode Materials for Low-Temperature Solid Oxide Fuel Cells: Fundamentals to Optimization. *Electrochemical Energy Reviews*. Springer June 1, 2022, pp 263–311. https://doi.org/10.1007/s41918-021-00098-3.

(7) Gao, Z.; Mogni, L. V.; Miller, E. C.; Railsback, J. G.; Barnett, S. A. A Perspective on Low-Temperature Solid Oxide Fuel Cells. *Energy and Environmental Science*. Royal Society of Chemistry May 1, 2016, pp 1602–1644. https://doi.org/10.1039/c5ee03858h.

(8) Vohs, J. M.; Gorte, R. J. High-Performance SOFC Cathodes Prepared by Infiltration. *Advanced Materials*. March 6, 2009, pp 943–956. https://doi.org/10.1002/adma.200802428.

(9) Cao, J.; Ji, Y.; Shao, Z. Perovskites for Protonic Ceramic Fuel Cells: A Review. *Energy and Environmental Science*. Royal Society of Chemistry April 7, 2022, pp 2200–2232. https://doi.org/10.1039/d2ee00132b.

(10) Merkle, R.; Hoedl, M. F.; Raimondi, G.; Zohourian, R.; Maier, J. Oxides with Mixed Protonic and Electronic Conductivity. *Annu Rev Mater Res* **2021**, *51*, 461–493. https://doi.org/10.1146/annurev-matsci-091819-010219.

(11) Han, N.; Ren, R.; Ma, M.; Xu, C.; Qiao, J.; Sun, W.; Sun, K.; Wang, Z. Sn and Y Co-Doped $BaCo_{0.6}Fe_{0.4}O_{3-\delta}$ Cathodes with Enhanced Oxygen Reduction Activity and $CO_2$ Tolerance for Solid Oxide Fuel Cells. *Chinese Chemical Letters* **2022**, *33* (5), 2658–2662. https://doi.org/10.1016/j.cclet.2021.09.100.

(12) Gao, Y.; Zhang, M.; Fu, M.; Hu, W.; Tong, H.; Tao, Z. A Comprehensive Review of Recent Progresses in Cathode Materials for Proton-Conducting SOFCs. *Energy Reviews* **2023**, *2* (3), 100038. https://doi.org/10.1016/j.enrev.2023.100038.

(13) Duan, C.; Hook, D.; Chen, Y.; Tong, J.; O'Hayre, R. Zr and Y Co-Doped Perovskite as a Stable, High Performance Cathode for Solid Oxide Fuel Cells Operating below 500°C. *Energy Environ Sci* **2017**, *10* (1), 176–182. https://doi.org/10.1039/c6ee01915c.

(14) Lee, Y.-L.; Kleis, J.; Rossmeisl, J.; Shao-Horn, Y.; Morgan, D. Prediction of Solid Oxide Fuel Cell Cathode Activity with First-Principles Descriptors. *Energy Environ Sci* **2011**, *4*, 3966–3970. https://doi.org/10.1039/c1ee02032c.

(15) Lee, Y. L.; Lee, D.; Wang, X. R.; Lee, H. N.; Morgan, D.; Shao-Horn, Y. Kinetics of Oxygen Surface Exchange on Epitaxial Ruddlesden-Popper Phases and Correlations to First-Principles Descriptors. *Journal of Physical Chemistry Letters* **2016**, *7* (2), 244–249. https://doi.org/10.1021/acs.jpclett.5b02423.

(16) Jacobs, R.; Mayeshiba, T.; Booske, J.; Morgan, D. Material Discovery and Design Principles for Stable, High Activity Perovskite Cathodes for Solid Oxide Fuel Cells. *Adv Energy Mater* **2018**, *8* (11), 1–27. https://doi.org/10.1002/aenm.201702708.

(17) Jacobs, R.; Hwang, J.; Hong, W.; Shao-Horn, Y.; Morgan, D. Assessing Correlations with Electronic Structure Descriptors for Perovskite Catalytic Performance. *Chemistry of Materials* **2019**, *31* (3), 785–797. https://doi.org/10.1021/acs.chemmater.8b03840.

(18) Giordano, L.; Akkiraju, K.; Jacobs, R.; Vivona, D.; Morgan, D.; Shao-Horn, Y. Electronic Structure-Based Descriptors for Oxide Properties and Functions. *Acc Chem Res* **2022**, *55* (3), 298–308.

(19) Mayeshiba, T. T.; Morgan, D. D. Factors Controlling Oxygen Migration Barriers in Perovskites. *Solid State Ion* **2016**, *296*, 71–77. https://doi.org/10.1016/j.ssi.2016.09.007.

(20) Jacobs, R.; Booske, J.; Morgan, D. Understanding and Controlling the Work Function of Perovskite Oxides Using Density Functional Theory. *Adv Funct Mater* **2016**, *26* (30). https://doi.org/10.1002/adfm.201600243.

(21) Ma, T.; Jacobs, R.; Booske, J.; Morgan, D. Discovery and Engineering of Low Work Function Perovskite Materials. *J Mater Chem C Mater* **2021**, *9* (37), 12778–12790. https://doi.org/https://doi.org/10.1039/D1TC01286J.

(22) Lee, Y.; Gadre, M. J.; Shao-horn, Y.; Morgan, D. Ab Initio GGA+U Study of Oxygen Evolution and Oxygen Reduction Electrocatalysis on the (001) Surfaces of Lanthanum Transition Metal Perovskites $LaBO_3$ (B = Cr, Mn, Fe, Co and Ni). *Physical Chemistry Chemical Physics* **2015**, *17*, 21643–21663. https://doi.org/10.1039/C5CP02834E.

(23) Grimaud, A.; May, K. J.; Carlton, C. E.; Lee, Y.-L.; Risch, M.; Hong, W. T.; Zhou, J.; Shao-Horn, Y. Double Perovskites as a Family of Highly Active Catalysts for Oxygen Evolution in Alkaline Solution. *Nat Commun* **2013**, *4*, 1–7. https://doi.org/10.1038/ncomms3439.

(24) Kim, Y.; Ha, M.; Anand, R.; Zafari, M.; Baik, J. M.; Park, H.; Lee, G. Unveiling a Surface Electronic Descriptor for Fe-Co Mixing Enhanced the Stability and Efficiency of Perovskite Oxygen Evolution Electrocatalysts. *ACS Catal* **2022**, 14698–14707. https://doi.org/10.1021/acscatal.2c04424.

(25) Schmidt, J.; Marques, M. R. G.; Botti, S.; Marques, M. A. L. Recent Advances and Applications of Machine Learning in Solid-State Materials Science. *NPJ Comput Mater* **2019**, *5* (1). https://doi.org/10.1038/s41524-019-0221-0.

(26) Morgan, D.; Jacobs, R. Opportunities and Challenges for Machine Learning in Materials Science. *Annu Rev Mater Res* **2020**, *50*, 71–103. https://doi.org/10.1146/annurev-matsci-070218-010015.

(27) Ramprasad, R.; Batra, R.; Pilania, G.; Mannodi-Kanakkithodi, A.; Kim, C. Machine Learning and Materials Informatics: Recent Applications and Prospects. *NPJ Comput Mater* **2017**, No. July. https://doi.org/10.1038/s41524-017-0056-5.

(28) Agrawal, A.; Choudhary, A. Perspective: Materials Informatics and Big Data: Realization of the "Fourth Paradigm" of Science in Materials Science. *APL Mater* **2016**, *4* (5). https://doi.org/10.1063/1.4946894.

(29) Agrawal, A.; Choudhary, A. Deep Materials Informatics: Applications of Deep Learning in Materials Science. *MRS Commun* **2019**, *9* (3), 779–792. https://doi.org/10.1557/mrc.2019.73.

(30) Chen, C.; Zuo, Y.; Ye, W.; Li, X.; Deng, Z.; Ong, S. P. A Critical Review of Machine Learning of Energy Materials. *Adv Energy Mater* **2020**, *10* (8). https://doi.org/10.1002/aenm.201903242.

(31) Chen, A.; Zhang, X.; Zhou, Z. Machine Learning: Accelerating Materials Development for Energy Storage and Conversion. *InfoMat* **2020**, *2* (3), 553–576. https://doi.org/10.1002/inf2.12094.

(32) Weng, B.; Song, Z.; Zhu, R.; Yan, Q.; Sun, Q.; Grice, C. G.; Yan, Y.; Yin, W.-J. Simple Descriptor Derived from Symbolic Regression Accelerating the Discovery of New Perovskite Catalysts. *Nat Commun* **2020**, *11* (1), 3513. https://doi.org/10.1038/s41467-020-17263-9.

(33) Lunger, J. R. ,; Karaguesian, J.; Chun, H.; Peng, J.; Tseo, Y.; Shan, C. H.; Han, B.; Shao-Horn, Y.; Gomez-Bombarelli, R. Atom-by-Atom Design of Metal Oxide Catalysts for the Oxygen Evolution Reaction with Machine Learning. *ArXiv* **2023**.

(34) Zhai, S.; Xie, H.; Cui, P.; Guan, D.; Wang, J.; Zhao, S.; Chen, B.; Song, Y.; Shao, Z.; Ni, M. A Combined Ionic Lewis Acid Descriptor and Machine-Learning Approach to Prediction of Efficient Oxygen Reduction Electrodes for Ceramic Fuel Cells. *Nat Energy* **2022**, *7* (9), 866–875. https://doi.org/10.1038/s41560-022-01098-3.

(35) Xu, P.; Lu, T.; Ji, X.; Li, M.; Lu, W. Machine Learning Combined with Weighted Voting Regression and Proactive Searching Progress to Discover $ABO_{3-\delta}$ Perovskites with High Oxide Ionic Conductivity. *The Journal of Physical Chemistry C* **2023**. https://doi.org/10.1021/acs.jpcc.3c02893.

(36) Zhang, Y.; Xu, X. Modeling Oxygen Ionic Conductivities of $ABO_3$ Perovskites through Machine Learning. *Chem Phys* **2022**, *558*, 111511. https://doi.org/10.1016/j.chemphys.2022.111511.

(37) Schlenz, H.; Baumann, S.; Meulenberg, W. A.; Guillon, O. The Development of New Perovskite-Type Oxygen Transport Membranes Using Machine Learning. *Crystals (Basel)* **2022**, *12* (7), 947. https://doi.org/10.3390/cryst12070947.

(38) Xin, H. Catalyst Design with Machine Learning. *Nat Energy* **2022**, *7* (9), 790–791. https://doi.org/10.1038/s41560-022-01112-8.

(39) Jacobs, R.; Liu, J.; Abernathy, H.; Morgan, D. A Critical Assessment of Electronic Structure Descriptors for Predicting Perovskite Catalytic Properties. *ArXiv* **2023**. https://doi.org/https://arxiv.org/abs/2310.17744.

(40) Jacobs, R.; Mayeshiba, T.; Afflerbach, B.; Miles, L.; Williams, M.; Turner, M.; Finkel, R.; Morgan, D. The Materials Simulation Toolkit for Machine Learning (MAST-ML): An Automated Open Source Toolkit to Accelerate Data-Driven Materials Research. *Comput Mater Sci* **2020**, *176*, 109544. https://doi.org/10.1016/j.commatsci.2020.109544.

(41) Palmer, G.; Du, S.; Politowicz, A.; Emory, J. P.; Yang, X.; Gautam, A.; Gupta, G.; Li, Z.; Jacobs, R.; Morgan, D. Calibration after Bootstrap for Accurate Uncertainty Quantification in Regression Models. *NPJ Comput Mater* **2022**, *8* (1), 1–9. https://doi.org/10.1038/s41524-022-00794-8.

(42) Lundberg, S. M.; Allen, P. G.; Lee, S.-I. A Unified Approach to Interpreting Model Predictions. In *31st Conference on Neural Information Processing Systems (NIPS)*; 2017.

(43) Xu, Y.; Xu, K.; Zhu, F.; He, F.; Zhang, H.; Fang, C.; Liu, Y.; Zhou, Y.; Choi, Y.; Chen, Y. A Low-Lewis-Acid-Strength Cation Cs+-Doped Double Perovskite for Fast and Durable Oxygen Reduction/Evolutions on Protonic Ceramic Cells. *ACS Energy Lett* **2023**, *8*, 4145–4155.

(44) Ong, S. P.; Davidson, W.; Jain, A.; Hautier, G.; Kocher, M.; Cholia, S.; Gunter, D.; Chevrier, V. L.; Persson, K. A.; Ceder, G. Python Materials Genomics ( Pymatgen ): A Robust , Open-Source Python Library for Materials Analysis. *Comput Mater Sci* **2013**, *68*, 314–319. https://doi.org/10.1016/j.commatsci.2012.10.028.

(45) Li, W.; Jacobs, R.; Morgan, D. Predicting the Thermodynamic Stability of Perovskite Oxides Using Machine Learning Models. *Comput Mater Sci* **2018**, *150*, 454–463. https://doi.org/10.1016/j.commatsci.2018.04.033.

(46) Data for "Machine Learning Design of Perovskite Catalytic Properties." https://doi.org/https://doi.org/10.6084/m9.figshare.24450445.v1.

(47) Pedregosa, F.; Varoquaux, G.; Gramfort, A.; Michel, V.; Thirion, B.; Grisel, O.; Blondel, M.; Prettenhofer, P.; Weiss, R.; Dubourg, V.; Vanderplas, J.; Passos, A.; Cournapeau, D.; Brucher, M.; Perrot, M.; Duchesnay, É. Scikit-Learn: Machine Learning in Python. *Journal of Machine Learning Research* **2011**, *12*.

(48) Chollet, F. *Keras*. https://github.com/keras-team/keras.

(49) Abadi, M.; Barham, P.; Chen, J.; Chen, Z.; Davis, A.; Dean, J.; Devin, M.; Ghemawat, S.; Irving, G.; Isard, M.; Kudlur, M.; Levenberg, J.; Monga, R.; Moore, S.; Murray, D. G.; Steiner, B.; Tucker, P.; Vasudevan, V.; Warden, P.; Wicke, M.; Yu, Y.; Zheng, X.; Brain, G. TensorFlow: A System for Large-Scale Machine Learning. In *12th USENIX Symposium on Operating Systems Design and Implementation (OSDI '16)*; 2016; pp 265–284. https://doi.org/10.1038/nn.3331.

(50) Ward, L.; Agrawal, A.; Choudhary, A.; Wolverton, C. A General-Purpose Machine Learning Framework for Predicting Properties of Inorganic Materials. *NPJ Comput Mater* **2016**, No. June, 1–7. https://doi.org/10.1038/npjcompumats.2016.28.

(51) Wu, H. H.; Lorenson, A.; Anderson, B.; Witteman, L.; Wu, H. H.; Meredig, B.; Morgan, D. Robust FCC Solute Diffusion Predictions from Ab-Initio Machine Learning Methods. *Comput Mater Sci* **2017**, *134*, 160–165. https://doi.org/10.1016/j.commatsci.2017.03.052.

(52) Lu, H. J.; Zou, N.; Jacobs, R.; Afflerbach, B.; Lu, X. G.; Morgan, D. Error Assessment and Optimal Cross-Validation Approaches in Machine Learning Applied to Impurity Diffusion. *Comput Mater Sci* **2019**, *169*, 109075. https://doi.org/10.1016/j.commatsci.2019.06.010.

(53) Jacobs, R. M.; Booske, J. H.; Morgan, D. Intrinsic Defects and Conduction Characteristics of Sc2O3 in Thermionic Cathode Systems. *Phys Rev B* **2012**, *86* (5), 054106.

(54) Lee, Y.-L.; Kleis, J.; Rossmeisl, J.; Morgan, D. Ab Initio Energetics of LaBO_{3}(001) ( B=Mn , Fe, Co, and Ni) for Solid Oxide Fuel Cell Cathodes. *Phys Rev B* **2009**, *80* (22). https://doi.org/10.1103/PhysRevB.80.224101.